\documentclass[fleqn,usenatbib]{mnras}

\usepackage{newtxtext,newtxmath}

\usepackage[T1]{fontenc}
\usepackage{ae,aecompl}

\usepackage{graphicx}	
\usepackage{amsmath}	
\usepackage{amssymb}	
\usepackage{color}      

\usepackage{hyperref}
\def\xlinkspace#1 #2{%
 \ifx\relax#2%
 \xlinkdash#1-\relax
 \else
 \xlinkdash#1 -\relax
 \expandafter\xlinkspace\expandafter#2%
 \fi}

\def\xlinkdash#1-#2{%
 \ifx\relax#2%
 \tmp{#1}%
 \else
 \tmp{#1-}%
 \expandafter\xlinkdash\expandafter#2%
 \fi}

\newcommand{\kms}{km~s$^{-1}$}
\newcommand{\ie}{{\it i.e.}}

\newcommand{\be}{\begin{equation}}
\newcommand{\ee}{\end{equation}}

\definecolor{todo}{RGB}{200,0,0}

\title[HI-MaNGA]{HI-MaNGA: HI Follow-up for the MaNGA Survey}

\author[]{Karen L. Masters$^{1,2}$, David V. Stark$^{3}$, Zachary J. Pace$^4$, Frederika Phipps$^{5,2}$, \newauthor Wiphu Rujopakarn$^{6,7,3}$, Nattida Samanso$^6$, Emily Harrington$^{8,1}$, \newauthor Jos{\'e} R. S{\'a}nchez-Gallego$^{9}$,  Vladimir Avila-Reese$^{10}$, Matthew Bershady$^{4,11}$, \newauthor Brian Cherinka$^{12}$,  Catherine E. Fielder$^{13}$, Daniel Finnegan$^{14,1}$,  Rogemar A. Riffel$^{15,16}$, \newauthor Kate Rowlands$^{13}$,  Shoaib Shamsi$^1$, Lucy Newnham$^{2,1}$,  Anne-Marie Weijmans$^{17}$,  \newauthor Catherine A. Witherspoon$^{4}$ \thanks{E-mail: klmasters@haverford.edu}
\\
$^{~1}$Department of Physics and Astronomy, Haverford College, 370 Lancaster Avenue, Haverford, Pennsylvania 19041, USA\\
$^{~2}$Institute of Cosmology \& Gravitation, University of Portsmouth, Dennis Sciama Building, Portsmouth, PO1 3FX, UK\\
$^{~3}$Kavli Institute for the Physics and Mathematics of the Universe, Todai Institutes for Advanced Study,  \\ The University of  Tokyo, Kashiwa, Japan 277- 8583\\
$^{~4}$Department of Astronomy, University of Wisconsin-Madison, 475 N. Charter St., Madison, Wisconsin 53726, USA\\
$^{~5}$School of Physics and Astronomy, University of Southampton, Southampton, SO17 1BJ, United Kingdom\\
$^{~6}$Department of Physics, Faculty of Science, Chulalongkorn University, 254 Phayathai Road, \\ Pathumwan,  Bangkok 10330, Thailand\\
$^{~7}$National Astronomical Research Institute of Thailand, Don Kaeo, Mae Rim, Chiang Mai 50180, Thailand\\
$^{~8}$Department of Physics, Bryn Mawr College, 101 N Merion Ave, Bryn Mawr, Pennsylvania 19010, USA\\
$^{9}$Department of Astronomy, Box 351580, University of Washington, Seattle, Washington 98195, USA\\
$^{10}$Instituto de Astronom{\'i}a, Universidad Nacional Aut\'onoma de M\'exico, A.P. 70-264, 04510, M\'exico, D.F., M\'exico\\
$^{11}$South African Astronomical Observatory, P.O. Box 9, Observatory 7935, Cape Town, South Africa\\
$^{12}$Department of Physics and Astronomy, Johns Hopkins University, 3400 N. Charles St., Baltimore, MD 21218, USA\\
$^{13}$PITT PACC, Department of Physics and Astronomy, University of Pittsburgh, Pittsburgh, PA 15260, USA\\
$^{14}$Department of Physics, Siena College, 515 Loudon Road, Loudonville, New York 12211, USA\\
$^{15}$Departamento de F{\'i}sica, CCNE, Universidade Federal de Santa Maria, 97105-900, Santa Maria, RS, Brazil\\
$^{16}$Laborat{\'o}rio Interinstitucional de e-Astronomia, 77 Rua General Jos{\'e} Cristino, Rio de Janeiro, 20921-400, Brazil\\
$^{17}$School of Physics and Astronomy, University of St Andrews, North Haugh, St Andrews, KY16 9SS, UK
}

\date{\today}

\pubyear{2018}

\begin{document}
\label{firstpage}
\pagerange{\pageref{firstpage}--\pageref{lastpage}}
\maketitle

\begin{abstract}
We present the HI-MaNGA programme of HI follow-up for the Mapping Nearby Galaxies at Apache Point Observatory (MaNGA) survey. MaNGA, which is part of the Fourth phase of the Sloan Digital Sky Surveys (SDSS-IV), is in the process of obtaining integral field unit (IFU) spectroscopy for a sample of $\sim 10,000$ nearby galaxies. We give an overview of the HI 21cm radio follow-up observing plans and progress and present data for the first 331 galaxies observed in the 2016 observing season at the Robert C. Bryd Green Bank Telescope (GBT). We also provide a cross match of the current MaNGA (DR15) sample with publicly available HI data from the Arecibo Legacy Fast Arecibo L-band Feed Array (ALFALFA) survey. The addition of HI data to the MaNGA data set will strengthen the survey's ability to address several of its key science goals that relate to the gas content of galaxies, while also increasing the legacy of this survey for all extragalactic science.
\end{abstract}


\section{Introduction}
 MaNGA (Mapping Nearby Galaxies at Apache Point Observatory; \citealt{Bundy2015}) is part of the SDSS-IV programme of surveys \citep{Blanton2017} which began in 2014 and is running until 2020. MaNGA modified the SDSS-III BOSS fibre fed spectrograph \citep{Smee2013} on the Sloan Foundation 2.5m telescope \citep{Gunn2006} to create pluggable Integral Field Units (IFUs) which group between 19-127 fibres in a hexagonal pattern (or ``bundle") across the face of each MaNGA galaxy \citep{Law2015}, ranging in size from 12-32\arcsec ~in diameter. This allows the survey to obtain spatially resolved spectra for a large sample of galaxies. The MaNGA instrument has seventeen such fibre bundles in each SDSS plate (a sky area with a diameter of 3$^\circ$). 
 
 MaNGA is observing $\sim$1600 galaxies/year for a planned sample of $\sim$10,000 galaxies over its full 6 year duration \citep{Law2015,Wake2017}. In the most recent public release (Data Release 15, or DR15, \citealt{DR15}) MaNGA data for 4,621 unique galaxies were made available to the community. These data already make MaNGA the largest IFU survey in the world (e.g. ATLAS-3D, CALIFA or SAMI with $N=260$, $N= 600$ and $N\sim3000$ respectively, \citealt{Cappellari2011,Sanchez2012,Bryant2015}), allowing the internal kinematics and spatially resolved properties of stellar populations and ionised gas to be studied as a function of local environment and halo mass across all types of galaxies.  

MaNGA will provide the most comprehensive census of the stellar (and ionised gas) content of local galaxies to-date, but galaxies are not made of stars alone. The science goals of MaNGA are focused on understanding the physical mechanisms which drive the evolution of the galaxy population. These goals have been developed into the four key science questions of MaNGA \citep{Bundy2015}, all of which crucially depend on understanding not only the stellar content but also the cold gas budget of galaxies in the MaNGA sample. In the next section we summarize how knowledge of HI content contributes to all of MaNGA's key science questions.

\subsection{How HI will Contribute to MaNGA Key Science Questions}
\begin{enumerate}
\item {\it How does gas accretion drive the growth of galaxies?} Information on the total cold gas content is a necessary first step to fully explore the role of gas accretion, by revealing the global HI content of each galaxy, and in particular galaxies found to have more HI than is typical may be used to reveal gas accretion. Asymmetry in the HI profile may also correlate with accretion \citep[e.g.][]{Bournaud2005}. Finally, knowledge of total content will also provide targets for spatially resolved HI follow-up to reveal the details of gas accretion.
\item {\it What are the relative roles of stellar accretion, major mergers, and instabilities in forming galactic bulges and ellipticals?} The cold gas content drives the dynamics of secular evolution (e.g. bars, \citealt{Athanassoula2003}), as dynamically cold gas is a more efficient transport of angular momentum than the stars. Modeling of the shape of the HI profile, combined with MaNGAs stellar and ionised gas velocity maps may allow us to statistically probe HI distributions - e.g. looking for central holes. This is a technique we plan to investigate in future work. Extended HI is also a better probe of interactions than stellar morphology \citep[e.g][]{Holwerda2011}. 
\item{\it What quenches star formation? What external forces affect star formation in groups and clusters?} Information about the cold gas content is crucial for understanding the physical mechanisms that regulate gas accretions and quench galaxy growth via the conversion of gas into stars \citep[e.g. see][who look at the links between AGN feedback and CO content]{Rosario2018}. HI-MaNGA data can be combined with CO follow-up to add information on the molecular hydrogen (e.g. ongoing CO followup surveys like MASCOT\footnote{\url{http://www.eso.org/~dwylezal/mascot}} and JINGLE, \citealt{Saintonge2018}\footnote{\url{http://www.star.ucl.ac.uk/JINGLE/}}) in order to complete this picture across a representative subset of the MaNGA sample. The efficiency of converting atomic into molecular hydrogen, given by the H$_2$-to-HI mass ratio, is tightly related to the large-scale star formation in galaxies \citep[e.g.,][]{Leroy+2008}.  Exploring the dependencies of this ratio on mass, mass surface density, galaxy type, specific star formation rate, and environment will help to clarify the role of global disk instabilities versus local processes of the ISM in the star formation efficiency of galaxies \citep[e.g.,][]{Blitz+2006,Krumholz+2009,Obreschkow+2009}. These can also be compared with the star formation histories (either from stellar population synthesis, or using current SFR via ionized gas) as well as metallicities obtained the MaNGA data, adding crucial information for this analysis. 
\item {\it How was angular momentum distributed among baryonic and non-baryonic components as the galaxy formed, and how do various mass components assemble and influence one another?} Without the full baryonic mass accounting for both stars and gas this question cannot be answered. Nowadays, the stellar-to-halo mass relation is one of the most used relations in extragalactic astronomy \citep[][]{Wechsler2018}. A generalization of it to the gaseous and total baryonic contents provides relevant information for understanding the galaxy-halo connection and the main physical processes that drive galaxy evolution. Volume weights can be applied to the MaNGA survey to produce a  volume-limited sample \citep[e.g.,][]{Wake2017}, in such a way that the galaxy-halo connection for stellar, HI, H$_2$, and baryonic masses will be possible. The baryonic Tully-Fisher relation \citep[e.g.][]{McGaugh2000,Stark2009,Avila-Reese2008} provides the most direct observational link between baryonic mass and dark halo mass. Molecular gas typically does not contribute significantly to the total baryonic mass ($M_{\rm H2} \sim 0.1$ M$_\star$; e.g. \citealt{Boselli2014}), but HI mass can be a significant fraction, or even the dominant component, in the mass range of the MaNGA sample and so the total HI mass must be directly measured. 
Further, MaNGA traces the stellar and ionised gas kinematics out to only 1.5$r_e$ or 2.5$r_e$ \citep{Law2015}. HI kinematics (rotation widths) will provide an anchor point for the rotation speed of galaxies in their outer parts.
\end{enumerate}

 In this paper we introduce HI-MaNGA, a program of HI (21cm line) followup of MaNGA galaxies aimed at contributing HI information to help MaNGA data be used to address its key science questions. This first HI-MaNGA paper is intended to introduce the survey and document the first release of data, which was released as a Value Added Catalogue (VAC) in SDSS-IV DR15  \citep{DR15}. We provide in this release data from our first year of observing at the Robert C. Byrd Green Bank Telescope (GBT; under project code AGBT16A\_095). This comprises the results of observations of 331 MaNGA galaxies. Observations have to-date been completed at GBT for a further $\sim$2000 MaNGA targets; those data will be released in the future. 
 
 The structure of this paper is as follows. We describe the target selection for HI-MaNGA and existing HI in \S \ref{sec:target}. Our observational strategy and data reduction process is described in \S \ref{sec:obs}. We show some overview results based on the HI content or dynamics of MaNGA galaxies in \S \ref{sec:result}, and conclude with a summary in \S \ref{sec:summary}

\section{Target Selection and Existing HI} \label{sec:target}
The basic selection for HI-MaNGA targets is all MaNGA observed galaxies with $cz<15,000$ \kms, and not obviously in the sky area observed by ALFALFA \citep{Haynes2011,Haynes2018}.\footnote{There is some deliberate overlap to check cross calibration. Also, as the final ALFALFA100 catalogue was not released at the start of HI-MaNGA there is some unintentional overlap at the edges of the surveys.}

\begin{figure}
\includegraphics[angle=0,width=7cm]{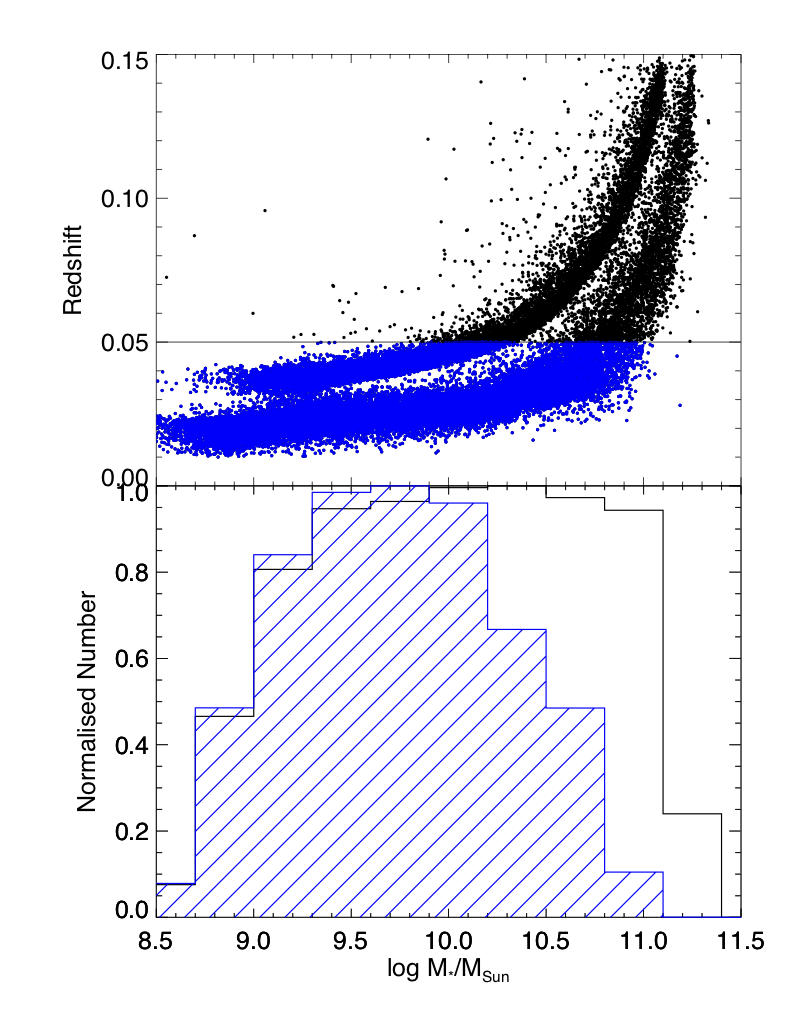}
\caption{We show the impact that a redshift limit of $cz<15,000$km/s which we apply to the HI-MaNGA follow-up programme has on the mass distribution of MaNGA targets. The upper panel shows the redshift-stellar mass distribution of all possible MaNGA targets (upper concentration of points shows the primary sample, lower the secondary). Indicated with the horizontal line is the redshift limit for HI-MaNGA. The lower panel shows the mass distribution for the full MaNGA (unfilled; showing roughly flat mass distribution for $M_\star=10^{9-11}$M$_\odot$) and HI-MaNGA target galaxies (blue hatched; basically MaNGA targets with $cz<15000$\kms).}
\label{fig:HImangamass}
\end{figure}

Our GBT observations (see \S \ref{sec:obs}) are designed to reach comparable $rms$ noise to the ALFALFA survey (around 1.5 mJy at 10 \kms ~velocity resolution \citealt{Haynes2011}); the upper redshift limit is chosen partly by the redshift range of ALFALFA, and partly by the distance at our expected depth where we expect more non-detections than detections. This redshift cut partially acts as a stellar mass limit in the MaNGA sample because of the way MaNGA is selected \citep{Wake2017}. We illustrate this in Figure \ref{fig:HImangamass} which shows the stellar mass redshift relation (upper) and the mass distribution (lower) of MaNGA and HI-MaNGA targets respectively. Notice how MaNGA has a flat mass distribution across $M_\star \sim 10^{9-11} M_\odot$, while HI-MaNGA targets are more strongly peaked at  $M_\star \sim 10^{9.8} M_\odot$: while basically all low mass MaNGA galaxies will be followed up in HI, higher mass galaxies are preferentially further away, and therefore less likely to be part of the follow-up presented here. By observing all MaNGA galaxies regardless of morphology we will provide an unbiased (or at least agnostic to morphological properties) census of the HI content and the impact this has on galaxy properties. 

\begin{figure*}
\includegraphics[angle=0,width=18cm]{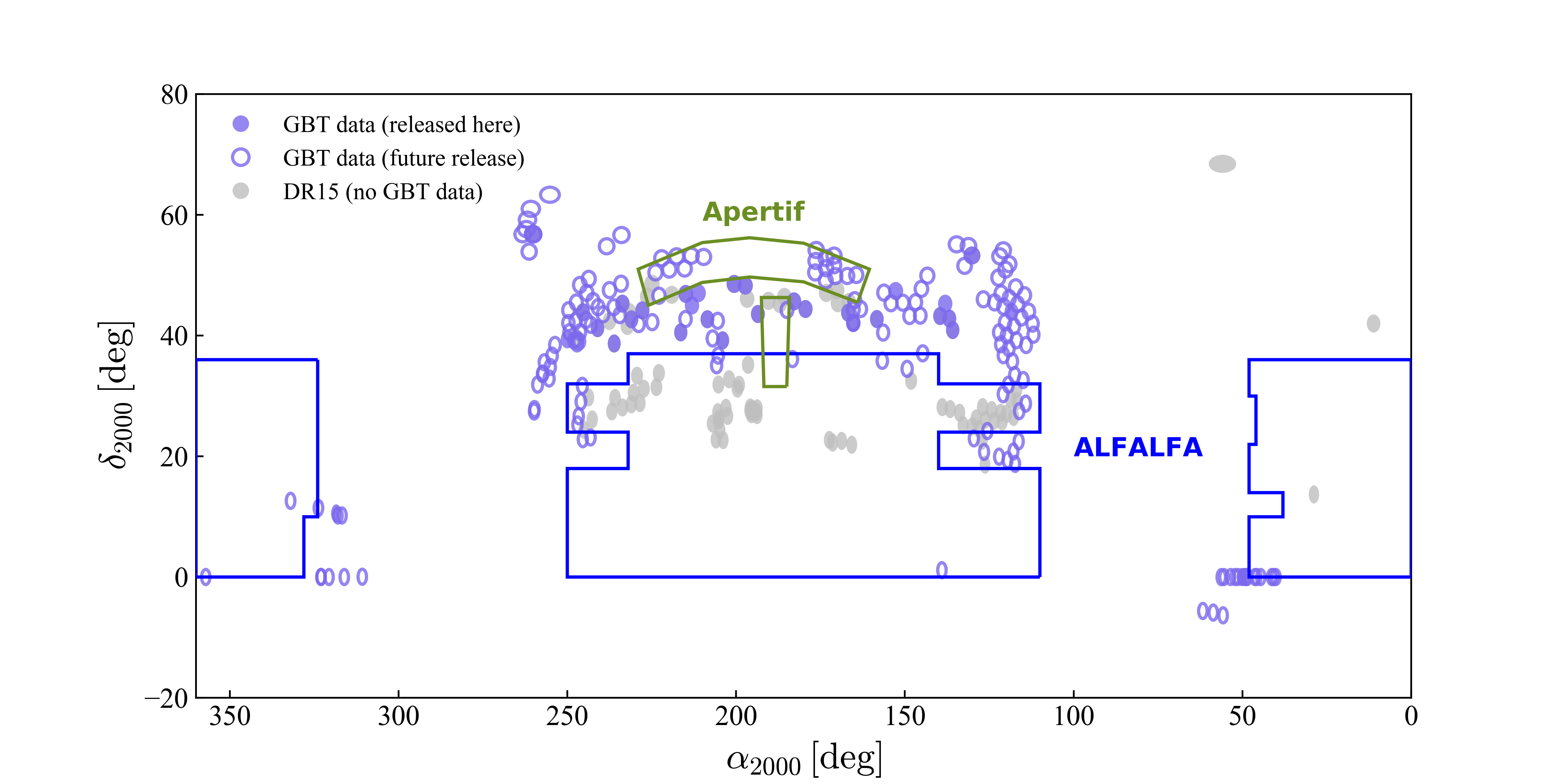}
\caption{The sky distribution of MaNGA observations and HI-MaNGA followup. The MaNGA DR15 sample is shown plotted as plates: in grey where there is no GBT data; open purple symbols where data has been taken, but not yet reduced; and filled purple circles show the sky positions of data released here. We also indicate the approximate footprint of the final ALFALFA survey \citep{Haynes2018} in blue and the planned Apertif medium deep survey.}
\label{fig:MaNGAsky}
\end{figure*}

 At the beginning of planning for HI-MaNGA, the MaNGA sample that was available was the ``MPL-4" list (``MaNGA Product Launch-4", an internal name for the subset of MaNGA observations which was then later released in DR13;  \citealt{DR13}). This means that all galaxies with HI measurements released in this preliminary data release are part of the DR13 (and therefore also DR15 and later) MaNGA samples. From within this list observing was completed in an order which maximised efficiency on sky, with a secondary goal of finishing HI observations for MaNGA galaxies on SDSS plates that were partially completed in earlier GBT observing sessions.
 We show in Figure \ref{fig:MaNGAsky} the sky distribution of MaNGA targets, observations and HI followup (as well as other relevant HI surveys).

As part of more recent planning for the HI-MaNGA observing, we also performed a cross-match of the MaNGA MPL-7 sample (the set which was released in DR15) with the final ALFALFA (100\%) release \citep{Haynes2018}. 
This provides all strong detections (roughly $S/N > 4.5$) in ALFALFA. We also extract upper limits for non-detections by measuring the noise at the sky and redshift position of each MaNGA galaxy directly from the cubes. We find a total of 908 of the MaNGA DR15 sample at $z<0.05$ in ALFALFA (334 detections and 574 upper limits).\footnote{Details of how to access this catalogue can be found at \url{https://www.sdss.org/dr15/data_access/value-added-catalogs/?vac_id=hi-manga-data-release-1}}

\section{GBT Observations and Data Reduction} \label{sec:obs}

In this paper we present observations from the first 331 HI-MaNGA targets, using 192.5 hours of GBT telescope time (or 35 minutes telescope time per galaxy). This was completed during the 2016A and 2016B observing semesters (all under proposal code AGBT16A\_95). 

\subsection{Observations}

Observations were performed using the L-band (1.15-1.73 GHz) receiver on GBT, which has a FWHM beam of 8.8' at these frequencies. We made use of the VErsatile GBT Astronomical Spectrometer (VEGAS) backend\footnote{For details on VEGAS see \url{http://www.gb.nrao.edu/vegas/report/URSI2011.pdf}}.  VEGAS was tuned to place 21cm (1420.405 MHz) emission at the known optical redshift of the MaNGA galaxy (from the NASA Sloan Atlas, \citealt{blanton2011}) at the centre of the bandpass, which was set to have a width of 23.44 MHz. A total of 4096 channels were used to collect data (which therefore had a raw spectral resolution of 5.72 kHz; or 1.2 km/s). As this is much smaller than needed to resolve the velocity structure of a typical galaxy, we boxcar smooth by a factor of four (to a resolution of 22.89 kHz, or $\sim$5.0 km/s) during the final data processing, and then performed a Hanning Smoothing for a final effective velocity resolution of 10 \kms.
\footnote{As galaxies in our sample range from $z=0.01 -0.05$ the exact value varies by about 5\% across the redshift range}

\begin{figure}
\includegraphics[angle=0,width=7.5cm]{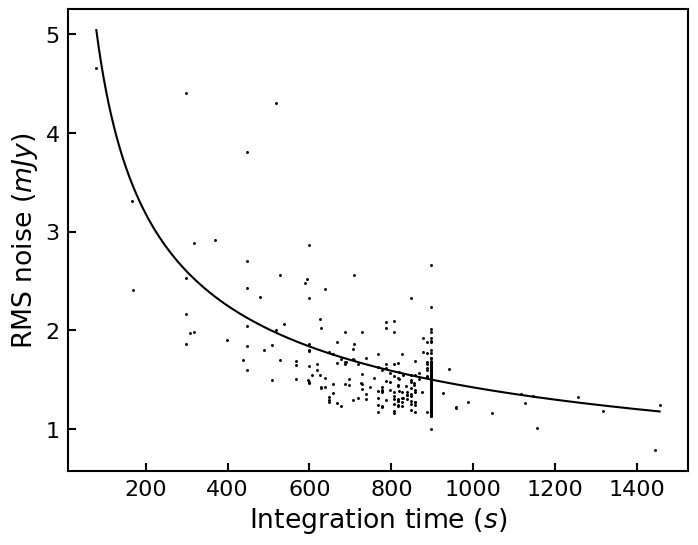}
\caption{We show the $rms$ noise as a function of integration time for our observing. The gathering of points at T = 900 s reveals our typical integration time around the targeted noise of 1.5 mJy. The solid line indicates a $t^{-1/2}$ relationship for 1.5 mJy in 900 s.}
\label{fig:noiseintergration}
\end{figure}

Observations were done in position switch mode using multiples of 5 min ON/OFF pairs (i.e. $\sim$10 mins telescope time). Data were collected in 10 second ``data samples" in order to mitigate the impact of time dependent radio frequency interference (RFI) causing catastrophic loss of entire samples (or more usually several samples in a row). In most cases each target was observed for a total of three ON/OFF pairs; sometimes, where a strong detection was found early observing this was cut short, and in some cases where significant interference from passing Global Positioning System (GPS) satellites ruined a significant fraction of ``samples" in an ON/OFF pair, an additional set (or sometimes more than one) was obtained. This procedure can be identified in Figure \ref{fig:noiseintergration} which shows the measured $rms$ noise as a function of total integration time in seconds. The vertical strip at $t=900$ s represents observations comprising three sets of 5 minute (or 300 s) ON/OFF pairs, while a large number of observations which lost small fractions of time to GPS or other interference scatter below or sometimes above this.

Figure \ref{fig:noiseintergration} also illustrates that our goal to obtain roughly $rms$=1.5 mJy observations has been largely achieved; where the noise is significantly higher this is typically because the galaxy was a strong HI emitter (and therefore detected even in a noisier spectrum). The solid line shows a behaviour of $t^{-1/2}$ normalized to 1.5 mJy at $t=900$ s. 

\subsection{Data Reduction} \label{sec:reduction}

Data was reduced making use of the custom GBTIDL\footnote{\url{http://gbtidl.nrao.edu/}} interface to IDL (the Interactive Data Language\footnote{\url{https://www.harrisgeospatial.com/SoftwareTechnology/IDL.aspx}}). Data segments free of GPS or other significant interference are first combined, edges trimmed, and narrow frequency RFI removed before smoothing to the final 10 \kms ~resolution. 

Calibration was performed using the GBT gain curves which are reported to be highly accurate at L-band for simple ON/OFF observing.\footnote{A flux scale accuracy of 10-20\% is reported in the GBTIDL Calibration Document at \url{http://wwwlocal.gb.nrao.edu/GBT/DA/gbtidl/gbtidl_calibration.pdf}} Finally baselines are fit to the signal free part of the spectrum. 

The reduced and baseline-fitted spectra for the first 331 targets observed at GBT on this programme are provided as a Value Added Catalogue in SDSS DR15 \citep{DR15} accessible on the SDSS Science Archive Server (SAS\footnote{\url{https://data.sdss.org/sas/mangawork/manga/HI/v1_0_1/spectra/GBT16A_095/}}); a detailed data model is provided.\footnote{\url{https://internal.sdss.org/dr15/datamodel/files/MANGA_HI/HIPVER/spectra/HIPROP/mangaHI.html}} For each observation we provide a row in an overview catalogue file,\footnote{\url{	https://data.sdss.org/sas/dr15/manga/HI/v1_0_1/mangaHIall.fits}} which also has a data model available.\footnote{ \url{https://internal.sdss.org/dr15/datamodel/files/MANGA_HI/HIPVER/mangaHIall.html}} This {\tt mangaHIall} file includes information on either the detection or non-detection as well as meta-data to aid in using in combination with MaNGA data. This is intended to be the structure for future larger data releases from the same program, which will have their own corresponding updated data models.  

It is also possible to access HI-MaNGA data using the Marvin interface \citep{Cherinka2018}.\footnote{For details on this see the tutorial at \url{https://sdss-marvin.readthedocs.io/en/stable/tools/catalogues.html\#value-added-catalogs-vacs}}

\subsubsection{Characterising Detections}\label{sec:detections}

As all galaxies are observed at their known optical redshift, we determine detection at a fixed smoothing scale by eye. This procedure is standard for similar single dish surveys; a more quantitative/automated detection scheme is being considered for future HI-MaNGA data releases. 

We report the peak $S/N$ calculated as $S/N = S_p/rms$. This will introduce a slight bias due to the measured $S_p$ being elevated by positive noise peaks. The user may prefer to re-calculate $S/N_c = (S_p - rms)/rms$ from tabulated values. The integrated $S/N$ is more appropriate to assess the significance of detections, and can be calculated as $S/N_{\rm int} = F_{\rm HI}/F_{\rm HI, error}$, where $F_{\rm HI, error}$ is described below.

\begin{figure*}
\includegraphics[angle=0,width=13.0cm]{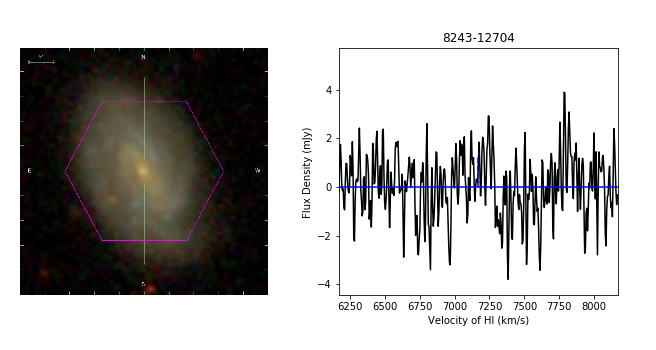}
\includegraphics[angle=0,width=13.0cm]{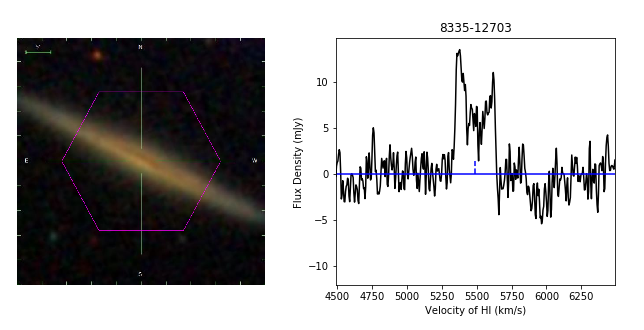}
\includegraphics[angle=0,width=13.0cm]{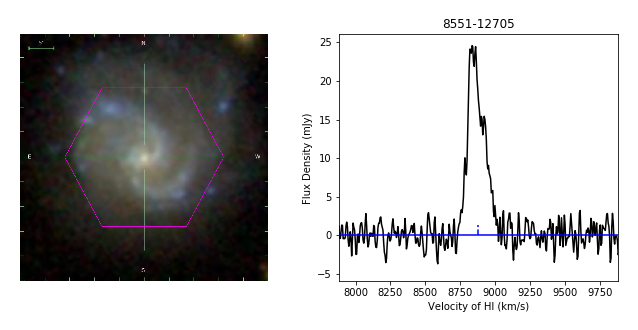}
\caption{Example spectra for three MaNGA galaxies with low (i.e. use with caution as it could be not real), average, and high S/N in the HI detection (peak $S/N$ values are 2.4, 7.5 and 17, while integrated $S/N$ using the flux error in Equation 1 are 2.9, 21 and 46 respectively). At the right is shown the baseline subtracted radio spectrum centered on the optical redshift of the galaxy (dashed line) whose SDSS $gri$ image is shown at left. The galaxies are (from top to bottom) MaNGAID=1-47291, 1-252072 and 1-247382. The MaNGA bundle is indicated by the purple hexagon; recall that the GBT beam at L-band is at least 18 times larger than this (8.8' compared to a maximum bundle size of 32.5").}
\label{fig:DetectionExample}
\end{figure*}

Example detections across low, median and high $S/N$ are shown in Figure \ref{fig:DetectionExample}. HI widths like these are characterised using the same procedure as was described in \citet{Masters2014}; based on \citet{Springob2005} this is also similar to the measurements performed by ALFALFA \citep{Haynes2018}. Not all measurements are possible on the lowest $S/N$ detections; which should always be used with caution as errors on extracted quantities will be large, and the likelihood of spurious detections is high. 

A summary of all measurements which are provided for each detection (where possible) is given in Table \ref{tab:measurements}. We refer the reader to \citet{Masters2014} and references therein for full details of these measurements, but provide here for convenience the formula used to calculate: 
\begin{enumerate}
\item The statistical error on the HI flux: 
\begin{equation}
F_{\rm HI,error} = rms \sqrt{\Delta v W},
\end{equation}
where $\Delta v = 10$\kms ~is the channel resolution (after Hanning smoothing), and $W$ should be the width of the profile (ideally the full width of baseline where signal is integrated, a value of $1.2 W_{\rm P20}$ can be used to approximate this). 
\item HI masses from fluxes: 
\be 
\label{eqn:HImass}
M_{\rm HI}/M_\odot = 2.356 \times 10^5 \left(\frac{D}{\rm Mpc}\right)^2 \left(\frac{F_{\rm HI}}{\rm Jy ~km/s}\right).
\ee
\end{enumerate}
We highlight that we provide raw widths and fluxes (and HI masses) in the catalogue. Users may wish to apply the following corrections to reconstruct more physically representative values: 
\begin{enumerate}
\item To correct HI masses for HI self-absorption you may like to use
\be
M_{\rm HI,c} = c M_{\rm HI}, 
\ee
where $c = (a/b)^{0.12}$ has been recommended (using the optical axial ratio $(a/b)$, see \citealt{Giovanelli1994} for details). 
\item To correct HI widths for inclination effects, cosmological broadening and the impact of turbulent motions and instrumental resolution use 
\begin{equation} \label{Wc}
W_c = \left[ \frac{W - 2\Delta v \lambda}{1+z} - \Delta t\right] \frac{1}{\sin i},
\end{equation}
with $\Delta v  = $ 5.00 \kms ~(the effective resolution before Hanning smoothing) and where $\lambda$ is a factor which accounts for the impact of noise on the effective resolution, taken from the simulations of \citet{Springob2005}.\footnote{We use the values for $\Delta v < 5$ \kms of $\lambda = 0.005$ for $\log(S/N)<0.6$, $\lambda=-0.4685+0.785\log(S/N)$ for $0.6<\log(S/N)<1.1$ and $\lambda=0.395$ for $\log(S/N)>1.1$.} The correction $\Delta t = 6.5 $ \kms ~is proposed to correct for turbulent motions, (also from the work of \citealt{Springob2005}) and the inclination $i$ can be calculated for a disc of intrinsic thickness, $q$ from its observed axial ratio $(a/b$) using
\be
\cos i =  \sqrt{\frac{(b/a)^2 - q^2}{1-q^2}},
\ee
and where $q=0.2$ is a reasonable average estimate for discs (see \citealt{Masters2014} and references therein). 
\item Cosmological corrections are small in this redshift range ($0.01<z<0.05$), however we list some here (and point the reader to \citealt{Meyer2017} for a full discussion). We re-iterate that these corrections have not been applied in our DR1 catalogue. 
\begin{itemize}
    \item The use of Jy \kms ~as units of flux (which is standard in HI surveys in the local Universe) introduces a $(1+z)^2$ term into the flux when expressed in units with the dimensions of flux (Jy Hz). This will propagate into all measurements using integrated flux (i.e. HI masses).
    \item Peculiar velocities can introduce significant distance errors in the local Universe \citep[e.g. as explored in][]{Masters2004}. However the minimum redshift limit of the MaNGA survey ($z>0.01$) means the impact of this is $< 10\%$ on HI-MaNGA masses.
    \item Widths are provided in rest frame. Equation \ref{Wc} includes the $(1+z)$ correction which should be applied to correct to observed frame.
\end{itemize}
\end{enumerate}

\begin{table*}
\caption{\label{tab:measurements} Summary of measurements made on HI detections}
\begin{tabular}{lll}
\hline
Name & Units & Description  \\
\hline
\hline
$S_p$ & mJy & The peak HI flux density. \\
$S/N$ & - & The peak signal to $rms$ noise ratio.  \\
$F_{\rm HI}$ & Jy \kms & The integrated HI flux. Note this is not self-absorption corrected. \\
$\log(M_{\rm HI}/M_\odot)$ &  - & Log of the HI mass (in solar masses) from Equation \ref{eqn:HImass} assuming $D = v_{\rm opt}/70$ km~s$^{-1}$~Mpc$^{-1}$ \\
& & and using the raw HI flux (no correction for self-absorption). \\
$V_{\rm HI}$ & \kms & Central redshift of the HI detection (using optical definition for redshift, and in the Barycentric frame).\\
$W_{M50}$ & \kms  & Width of the HI line measured at 50\% of the median (which is also the mean) of the two peaks.\\
$W_{P50}$ & \kms  & Width of the HI line measured at 50\% of the peak.\\
$W_{P20}$ & \kms  & Width of the HI line measured at 20\% of the peak.\\
$W_{2P50}$ & \kms  & Width of the HI line measured at 50\% of the peak on either side.\\
$W_{F50}$ & \kms  & Width of the HI line measured at 50\% of the peak$-rms$ on fits to the sides of the profile.\\
$P_r$, $P_l$  & mJy & The peak HI flux densities in the low and high velocity peaks respectively. \\ 
$a_r$, $a_l$ & mJy & Fit parameters in $F(v) = a + b v $ fits to either side of the profile (used in measuring $W_{F50}$),  \\
$b_r$, $b_l$ & mJy/(\kms)  & ~~~~ where the zeropoint of the velocity axis in the fit is defined as the central velocity of the HI.\\
\hline
\end{tabular}
\end{table*}

\subsubsection{Characterising Non-Detections}

Non-detections are reported just as the $rms$ noise across the spectrum (in mJy), but we also report a conservative estimate of the HI mass upper limit, assuming width of $W = 200$ \kms  ~to allow to calculate an estimate of the HI flux which could have remained undetected (to $1 \sigma$) as: 
\be 
F_{\rm HI,lim} < 200 ~rms ~{\rm mJy ~km s}^{-1}, 
\ee
and therefore the HI upper limit as 
\be 
M_{\rm HI,lim}/M_\odot < 2.356 \times 10^5 \left(\frac{D}{\rm Mpc}\right)^2 \left(\frac{F_{\rm HI,lim}}{\rm Jy ~km/s}\right),
\ee
assuming $D = v_{\rm opt}/70$ km s$^{-1}$ Mpc$^{-1}$ (where $v_{\rm opt}$ is the optical redshift of the MaNGA galaxies in the NSA). To be used for statistical analysis, this simple estimate should be corrected so it does not depend on the channel width of the observations (which is implicit in the measurement of the $rms$). A better choice of a 3-$\sigma$ upper limit (which we do not provide in this catalogue release, but which can be calculated from the information given) would be
\be 
F_{\rm HI, lim} = 3 ~rms ~ \sqrt{W \Delta v} ~~{\rm mJy~km~s}^{-1},
\label{eqn:limit}
\ee
where $\Delta v = 10$\kms ~is the velocity resolution (after Hanning smoothing), and $W$ is the assumed width (e.g. 200 \kms ~as used above, or this could be based on the optically measured rotation from MaNGA). Although channel size, $\Delta v$, is included in Eqn \ref{eqn:limit}, this calculated upper limit will not scale with channel size, as any increase/decrease in channel size will be canceled by a decrease/increase in $rms$ (which should be calculated at $\Delta v$ resolution). On average we find that Equation \ref{eqn:limit} gives an upper limit $\sim 1.5\times$ smaller than that we report in the catalogue (which can therefore be considered a more conservative upper limit) and should be more appropriate in terms of noise statistics. 

\section{Results} \label{sec:result}

The simplest result we can show is the detection fraction for the programme. This is summarized in Table \ref{tab:summary}. Out of 331 galaxies observed we report detections consistent with HI coming from the target galaxy in redshift in 181 cases (i.e. a detection fraction of 55\%). We further report 38 ``bonus'' detections,\footnote{\url{https://data.sdss.org/sas/dr15/manga/HI/v1_0_1/mangaHIbonus.fits}} representing HI detected either at a redshift significantly offset from the target, or in the OFF position. These results should be used with extreme caution as the object emitting the HI is unlikely to be centred in the GBT beam, and therefore beam attenuation may be significant.

\begin{table}
\caption{\label{tab:summary} Summary of first year of observing for HI-MaNGA at GBT (AGBT16A\_95)}
\begin{tabular}{lr}
\hline
Status & N$_{\rm galaxies}$  \\
\hline
\hline
All observed & 331  \\
Detections & 181  \\
Upper limits & 150 \\
Bonus detections & 38 \\
\hline
\end{tabular}

\end{table}

\begin{figure}
\includegraphics[angle=0,width=7cm]{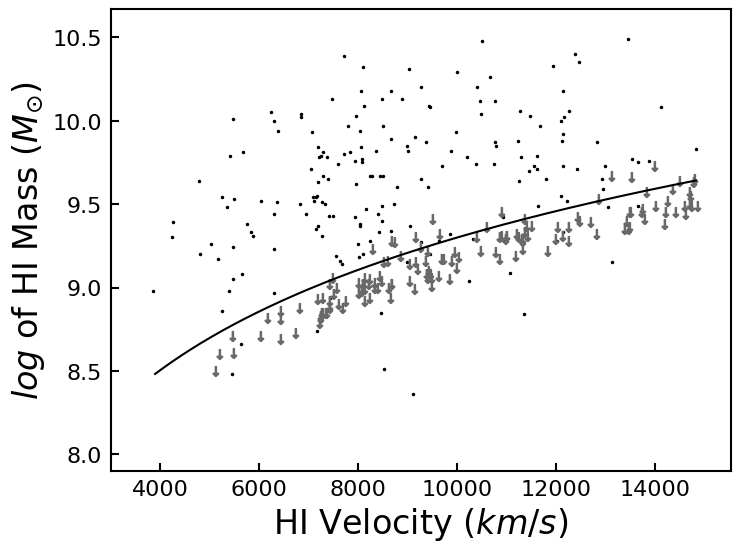}
\caption{We show the $\log$ HI mass (uncorrected) vs. HI recessional velocity for all GBT detections and non-detections released in this publication. The gray arrows indicate non-detections with the solid line being the upper limit of HI mass for these non-detections. The limit is derived from the inverse square relationship of mass and distance via our median value of $M_{\rm HI} = 10^{9.4} M_\odot$ being detectable at $cz = 9000$km/s}
\label{fig:HImasslimit}
\end{figure}

For all primary detections (and upper limits for the 150 non-detections), we show the HI mass (or limit) plotted against redshift in Figure \ref{fig:HImasslimit}. The solid line shows our estimated detection limit of $10^{9.4} M_\odot$ at a recessional velocity of $v= 9,000$ \kms ~(or a distance of 129 Mpc$/h_{70}$). There is some scatter around this line for observations with significantly higher or lower noise than typical (see Figure \ref{fig:noiseintergration} which shows the $rms$ noise of all observations). 

\subsection{HI Mass Fraction}

\begin{figure*}
\includegraphics[angle=0,width=16cm]{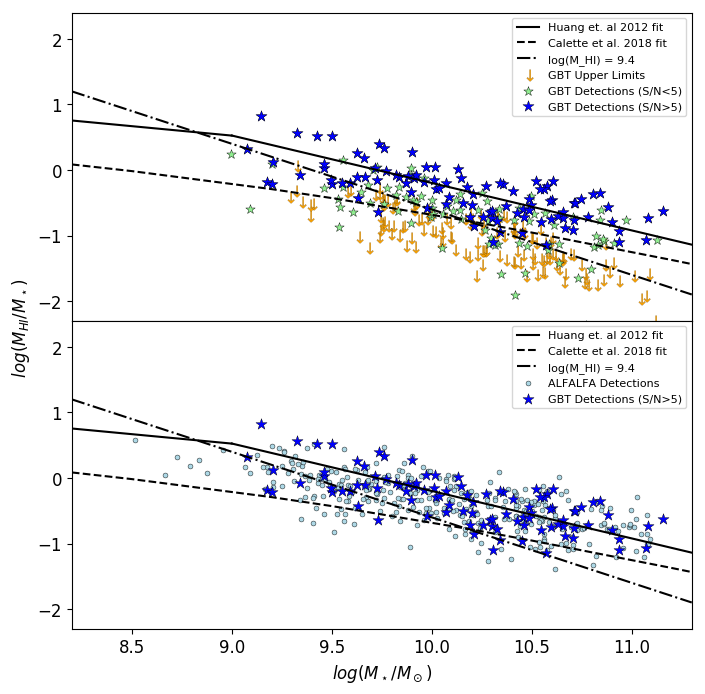}
\caption{The corrected HI mass fraction ($\log M_{\rm HI}/M_\star$) plotted against Pipe3D stellar masses for MaNGA galaxies. Upper: showing only data from the GBT observing published here. Lower: GBT strong detections plus ALFALFA data for MaNGA galaxies. The relations found by \citet{Huang2012} and \citet{Calette+2018} are overplotted as the solid and dashed lines respectively, while the dot-dashed line shows gas fraction for a constant HI mass of $\log{M_{\rm HI}/M_\odot} = 9.4$. \label{fig:HImassfraction} }
\end{figure*}

As a check on data quality, we plot in Figure \ref{fig:HImassfraction} our corrected HI mass fraction against stellar mass, and compare to results from ALFALFA matches to MaNGA galaxies, as well as the published relations based on all ALFALFA detections from \citet{Huang2012}, and the fit to a compilation and homogenization of data from various sources for late-type galaxies in \citet{Calette+2018}. We use stellar masses from the Pipe3D analysis tool \citep{sanchez2016a,sanchez2016b} applied to the MaNGA
data and presented in a Value Added Catalog \citep{sanchez2018}; here we use specifically the MPL-6 version of Pipe3d which used the same set of galaxies as released in DR15, but an earlier reduction pipeline. The HI masses here are corrected for self-absorption following the procedure in \S \ref{sec:detections}. Our results follow the published relation (and ALFALFA measurements) well, with some scatter to lower mass fractions, which are mostly low $S/N$ detections, and reflect the survey strategy as a follow-up to optical detections, rather than a blind HI survey like ALFALFA, which naturally picks up higher HI mass fraction galaxies in a stellar mass selected sample (because galaxies which scatter below the relation will preferentially have low $S/N$ detections which may not be believed in the targeted follow-up but not in a blind survey).

\subsection{Star Formation and HI Detections}

\begin{figure}
\includegraphics[angle=0,width=8.5cm]{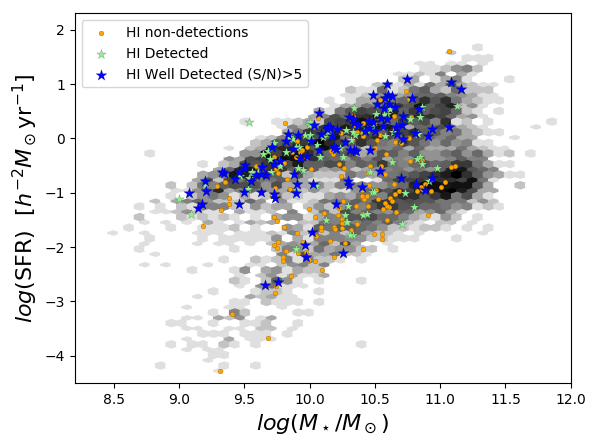}
\caption{Total star formation rate from the Pipe3D analysis of MaNGA data is plotted against the stellar mass of MaNGA galaxies. The entire DR15 MaNGA sample is shown in the greyscale contours (hexbin log scale with number), while those detected in HI are shown by the blue ($S/N<5$) or cyan stars ($S/N>5$), and non-detections are shown as red points. Note that the plotted HI data covers only a subset of DR15 galaxies. Never-the-less it's clear that while HI detections concentrate on the star forming sequence, they are not completely absent in quiescent galaxies.}
\label{fig:starformation}
\end{figure}

In Figure \ref{fig:starformation} we show a star formation stellar mass plot for the MaNGA DR15 sample. The integrated star-formation rates and stellar masses shown in this plot are taken from the Pipe3D analysis of MaNGA data \citep{sanchez2018}. All DR15 galaxies are shown in grey to reveal the typical distribution of MaNGA galaxies on the plot (with star forming galaxies in the upper sequence, and ``quiescent" galaxies below. We highlight HI non-detections (red points), weak detections (blue stars; $S/N<5$ in HI) and strong detections (cyan stars; $S/N>5$ in HI) from the HI data released with this publication, which we note does not cover all DR15 MaNGA galaxies (\ie a grey point means that the galaxy does not have HI data, not that it does not have HI). 

 As is expected, HI detections concentrate in the star forming sequence of this plot, however we note that detections are found in some quiescent MaNGA galaxies and some star forming galaxies have no detected HI. This trend has been previously noted in HI surveys \citep[e.g.][]{brown2015,saintonge2017}, who note that the molecular gas is more strongly correlated to the starformation properties than HI. Further work using this sample will investigate how the HI content of MaNGA galaxies correlates with star formation properties in more detail. 

\section{Summary and Conclusions} \label{sec:summary}

In this paper, we introduce the HI-MaNGA follow-up survey of the MaNGA sample \citep{Bundy2015}. This programme is aiming to obtain HI follow-up observations for a large subset of the MaNGA galaxies, selected only on redshift ($cz<15,000$ \kms). We present here the observational and data reduction strategy, as well as basic results from the first year of observing at the GBT (under project code AGBT16A\_95) which obtained HI measurements (or upper limits) for 331 MaNGA galaxies. These data are released as a VAC in SDSS DR15 \citep{DR15} available to download via \url{https://data.sdss.org/home} and with a catalogue available in CasJobs.\footnote{https://skyserver.sdss.org/CasJobs/}

These data are already in use by the wider MaNGA science team. Published work which has already made use of these GBT HI data include a study of the properties of quiescent dwarf galaxies \citep{Penny2016}, a paper on an unusual galaxy showing evidence for hot ionised gas infall (which is not detected in HI with GBT; \citealt{Lin2017Totoro}) and a paper which presents ALMA data for a sample of three green valley galaxies \citep{Lin2017ALMA}. 

We have performed a cross match of the MaNGA DR15 sample with the ALFALFA100 catalogue. We find 1308 of the MaNGA DR15 galaxies have HI data in ALFALFA (334 detections, and 574 upper limits). We provide our cross match as an electronic table. 

 We show some simple plots using these data in combination with MaNGA measurements (or other ancillary data). These include the HI mass fraction as a function of stellar mass, and an illustration of where HI detections lie on the star-formation--stellar mass plot.
 These 
 provide an illustration of the kind of science which will be enabled by HI follow-up for MaNGA. 

These data will provide a valuable resource to  combine with MaNGA data for studies of galaxy evolution and understanding the role of cold gas content which we will explore in future work. The addition of HI data to the MaNGA data set will strengthen the survey's ability to address several of its key science goals that relate to the gas content of galaxies, while also increasing the legacy of this survey for all extragalactic science.


\paragraph*{ACKNOWLEDGEMENTS.} 

The Green Bank Observatory is a facility of the National Science Foundation operated under cooperative agreement by Associated Universities, Inc. We would like to acknowledge the many GBT operators who helped implement this programme, which was entirely conducted with remote observations. 

Funding for the Sloan Digital Sky Survey IV has been provided by the Alfred P. Sloan Foundation, the U.S. Department of Energy Office of Science, and the Participating Institutions. SDSS-IV acknowledges
support and resources from the Center for High-Performance Computing at
the University of Utah. The SDSS web site is www.sdss.org.

SDSS-IV is managed by the Astrophysical Research Consortium for the Participating Institutions of the SDSS Collaboration including the Brazilian Participation Group, the Carnegie Institution for Science, Carnegie Mellon University, the Chilean Participation Group, the French Participation Group, Harvard-Smithsonian Center for Astrophysics, Instituto de Astrof\'isica de Canarias, The Johns Hopkins University, Kavli Institute for the Physics and Mathematics of the Universe (IPMU) / University of Tokyo, Korean Participation Group, Lawrence Berkeley National Laboratory, Leibniz Institut f\"ur Astrophysik Potsdam (AIP), Max-Planck-Institut f\"ur Astronomie (MPIA Heidelberg), Max-Planck-Institut f\"ur Astrophysik (MPA Garching), Max-Planck-Institut f\"ur Extraterrestrische Physik (MPE), National Astronomical Observatories of China, New Mexico State University, New York University, University of Notre Dame, Observat\'ario Nacional / MCTI, The Ohio State University, Pennsylvania State University, Shanghai Astronomical Observatory, United Kingdom Participation Group, Universidad Nacional Aut\'onoma de M\'exico, University of Arizona, University of Colorado Boulder, University of Oxford, University of Portsmouth, University of Utah, University of Virginia, University of Washington, University of Wisconsin, Vanderbilt University, and Yale University.

This work makes use of the ALFALFA survey, based on observations made with the Arecibo Observatory. The Arecibo Observatory is operated by SRI International under a cooperative agreement with the National Science Foundation (AST-1100968), and in alliance with Ana G. M\'endez-Universidad Metropolitana, and the Universities Space Research Association. We wish to acknowledge all members of the ALFALFA team for their work in making ALFALFA possible, and we also thank Martha Haynes for granting access to ALFALFA cubes to calculate upper limits for all MaNGA galaxies. 

This research made use of Marvin, a core Python package and web framework for MaNGA data, developed by Brian Cherinka, Jos\'e S\'anchez-Gallego, Brett Andrews, and Joel Brownstein. \citep{Cherinka2018}.

W.R. is supported by the Thailand Research Fund/Office of the Higher Education Commission Grant Number MRG6080294 and Chulalongkorn University's CUniverse.

FP and DF acknowledge Summer Research Funding from the South East Physics Network (www.sepnet.ac.uk) and Keck Northeast Astronomy Consortium (KNAC) respectively.

MAB acknowledges NSF Award AST-1517006.
 
We acknowledge the careful reading of the anonymous referee who helped catch many small (and not so small) errors in the first version of this paper.


\bibliography{references}{}
\bibliographystyle{apj}

\label{lastpage}
\end{document}